# Landscape of competing stripe and magnetic phases in cuprates

Yubo Zhang,[1] Christopher Lane,[2] James W. Furness,[1] Bernardo Barbiellini,[3,2] Robert S. Markiewicz,[2] Arun Bansil,[2,*] and Jianwei Sun[1,*]

[1]*Department of Physics and Engineering Physics, Tulane University, New Orleans, LA 70118, USA.*

[2]*Department of Physics, Northeastern University, Boston, MA 02115, USA.*

[3]*Department of Physics, School of Engineering Science, Lappeenranta University of Technology, FI-53851 Lappeenranta, Finland.*

*Correspondence to: jsun@tulane.edu and ar.bansil@neu.edu

Realistic modeling of competing phases in complex quantum materials has proven extremely challenging. For example, much of the existing density-functional-theory-based first-principles framework fails in the cuprate superconductors. Various many-body approaches involve generic model Hamiltonians and do not account for the couplings between spin, charge, and lattice. Here, by deploying the recently constructed strongly-constrained-and-appropriately-normed density functional, we show how landscapes of competing stripe and magnetic phases can be addressed on a first-principles basis in $YBa_2Cu_3O_6$ and $YBa_2Cu_3O_7$ as archetype cuprate compounds. We invoke no free parameters such as the Hubbard $U$, which has been the basis of much of the cuprate literature. Lattice degrees of freedom are found to be crucially important in stabilizing the various phases.

Competing orders lie at the heart of myriad fascinating properties of complex materials and their evolution with external controls of temperature, pressure, doping and magnetic field. The half-filled parent compounds of the cuprates, for example, are anti-ferromagnetic (AFM) insulators, which become high-temperature superconductors when doped with holes or electrons. This insulator-superconductor transformation is very complex and involves the presence of an intervening pseudogap phase and mechanisms for arresting superconductivity with increasing doping, so that the superconducting phase occupies a characteristic dome-shaped region in the phase diagram. Along these lines, a wide variety of orders and the associated phase diagrams are exhibited by iron-based superconductors, heavy-fermion compounds, organics, iridates, among other correlated materials of current interest.

Complex materials, which often involve strong electronic correlations, present a challenge to first-principles approaches. For example, the local-spin-density and the generalized-gradient approximations used commonly within the first-principles density functional theory (DFT) framework, yield a metallic rather than the experimentally observed insulating AFM ground state in undoped cuprates, so that a meaningful treatment of doping-dependent electronic structures of cuprates on this basis becomes impossible. The AFM state can be stabilized by invoking an *ad* *hoc* Hubbard *U* parameter, but that limits the predictive power of the theory. An alternate route that has been pursued is to deploy effective model Hamiltonians and attempt an exact treatment of electron interaction effects using quantum Monte Carlo (QMC) and other techniques. However, in view of their heavy computational cost, such studies have to be limited to fairly small clusters and one-band or at most three-band models in the cuprates, and cannot be material-specific or allow modifications in the Hamiltonian in response to interactions between the charge, spin and lattice degrees of freedom in the system.

Recent progress in constructing advanced density-functionals offers a new pathway for addressing at the first-principles level the electronic structures of correlated materials. In particular, the strongly-constrained-and-appropriately-normed (SCAN) functional(*1,2*) has been shown to provide a viable parameter-free description of the half-filled AFM ground state of the cuprates and their transition from the insulator to the metallic state with doping.(*3*) In $La_2CuO_4$, in this way, one correctly captures the size of the optical band gap, value of the copper magnetic moment and its alignment in the cuprate plane, and the magnetic form factor in good accord with the corresponding experimental results.(*4*)

Here, we discuss energies and other characteristics of magnetic and stripe phases in $YBa_2Cu_3O_6$ ($YBCO_6$) and $YBa_2Cu_3O_7$ ($YBCO_7$), as exemplar cuprate compounds, using SCAN-based DFT computations. $YBCO_6$ is representative of the half-filled parent insulating compounds, while $YBCO_7$ lies deep within the superconducting dome around the optimally doped region in the phase diagram. No free parameters are involved. In particular, the Hubbard *U* parameter is not invoked. All computations are carried to a high degree of self-consistency, and fully account for interactions between the spin, charge, and lattice degrees of freedom. 20 distinct, stripe phases are identified in $YBCO_7$ and 6 in $YBCO_6$. The ground state in $YBCO_6$ is reasonably well separated from the nearest stripe state by ~10 meV/planar-Cu (120 K). In sharp contrast, in $YBCO_7$, a large number of states are found to be nearly degenerate with the ground state. All stripe phases in $YBCO_7$ ($YBCO_6$) carry persistent copper magnetic moments of about 0.4 (0.5) $\mu_B$/Cu. Our analysis indicates that the spin, charge, and lattice degrees of freedom must be treated on an equal footing. Lattice degrees of freedom, for example, play a key role in stabilizing the stripe phases, and the self-consistency process between the various degrees of freedom induces substantial changes in stripe periodicity and the details of spin and charge distributions.

A stripe state is a special type of combination of spin- and charge-density wave orders in which domains of approximately $(\pi,\pi)$ AFM order are separated by antiphase boundaries (APBs) with reduced magnetic moments and excess charge density.(*5-12*) Fig. 1(a) shows a schematic one-dimensional (1D) stripe with three unit width, where a unit refers to a $CuO_2$ square or equivalently a Cu atom. The charge density in this stripe has a periodicity of $P_c = 3$; the spin-density or magnetism also has a periodicity of $P_m = 3$. This is a so-called bond-centered stripe because the APB passes through the center of Cu-Cu bonds. The central portion of the figure contains excess charge density (denoted by the large size of the green filled circles) and thus constitutes a charge-stripe. In contrast, the outer portions of the figure contain excess spin-density (denoted by large pink arrows) and thus makes a magnetic-stripe. Fig. 1(b) shows a schematic of a stripe of $P_c = 4$, but now the magnetic periodicity $P_m = 8$. Note that this is a site-centered stripe since the APB now passes through the Cu sites. Due to the presence of the APB, $P_m = 2P_c$ for even values $P_c$, but when $P_c$ is odd, $P_m = P_c$.

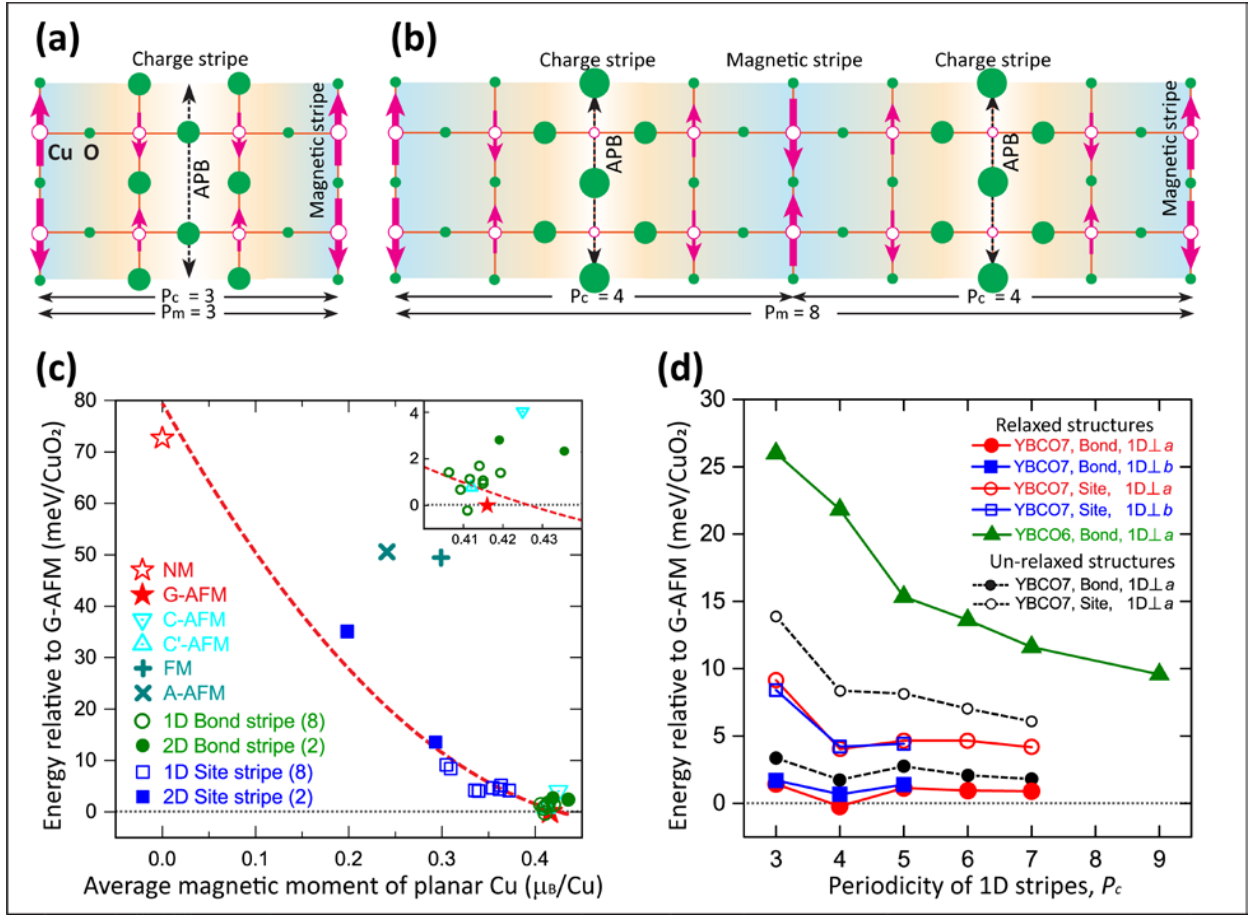

**Fig. 1. Schematics of bond-centered and site-centered stripes and the stability of our various computed phases in YBCO$_6$ and YBCO$_7$.** (a) Schematic of a bond-centered stripe in which the antiphase boundary (APB) passes through oxygen atoms (green dots) that lie in-between copper (red circles) atoms; $P_c = P_m = 3$, where $P_c$ and $P_m$ denote periodicities (in terms of the number of Cu atoms or CuO$_2$ squares) of the charge and spin densities, respectively. (b) Schematic of a site-centered stripe in which the APB passes through copper atoms with $P_c = 4$ and $P_m = 8$. (c) Energies of YBCO$_7$ stripe structures relative to the ground-state G-AFM structure as a function of the average magnetic moment of planar Cu atoms. The dashed red line is a guide to the eye. See Supplementary Materials for details of the various structures. G-AFM, C-AFM, C'-AFM, FM, and A-AFM are uniform or non-stripe states. The integers in parenthesis in the legend give the total number of different structures that are presented in the figure for various types of stripes (e.g., 8 different green circles are shown for 1D bond-type stripes). Inset is a blow-up of the lower right corner of the plot. (d) Relative energies with respect to the G-AFM state as a function of the periodicity of the various 1D charge stripes (perpendicular to the *a*- or *b*-axis as

indicated in the legend) in YBCO$_7$ and YBCO$_6$. Results for two un-relaxed stripe structures based on the G-AFM structure are given to highlight the substantial effects induced via lattice relaxation.

Figure 1(c) compares the relative energies of 26 different states we have found in YBCO$_7$. The energy zero is the energy of the ground state G-AFM structure. Both 1D and 2D stripe states are considered along with a number of uniform or non-stripe states (G-AFM, C-AFM, C'-AFM, FM, and A-AFM). Details of atomic structures, energies, and other features of all these predicted states are given in Sections 2 and 3 of the Supplementary Materials. Many states are nearly degenerate with the ground state with energy differences of a few meVs, see inset in Fig. 1(c). These results suggest a compelling model of the pseudogap phase, represented here by YBCO$_7$, in which many competing phases must be present, although the role of the associated magnetic fluctuations in creating the superconducting glue remains to be ascertained.

Interestingly, a simple organizing principle emerges for the magnetic order: the red-dashed trend line shows that the stability of magnetic states increases with increasing average magnetic moment on the planar copper atoms, small-scale deviations seen in the inset notwithstanding. This result is consistent with Mott physics in the sense that a Mott insulator is characterized by restricting hopping on or off of a particular atom to avoid double occupancy and to thus promote larger magnetic moments. In keeping with this trend, we can rule out any role of the non-magnetic Fermi-liquid state [NM, red unfilled star in Fig. 3(c)] in the low-energy physics of YBCO$_7$ because the non-magnetic state lies at a very high energy.

Figure 1(d) focuses on delineating how the energies of various 1D stripe phases evolve with increasing charge periodicity $P_c$. The difference between YBCO$_6$ and YBCO$_7$ is striking. While YBCO$_7$ exhibits a multitude of low-lying competing states, in sharp contrast, YBCO$_6$ possesses a well-defined G-AFM ground state (energy zero in the figure, dotted horizontal line) with no nearby competing states. The precursor stripes in YBCO$_6$ lie at quite high energies (filled green triangles), and are seen to be well-separated from the complex of low-energy states in YBCO$_7$. These results give some insight into nematicity effects, which refer to enhanced differences in properties along the $a$- and $b$-axis, that have been observed experimentally in many cuprates. As discussed in Supplementary Materials Section 3.2, an analysis of our results is consistent with the experimental finding that nematicity effects are stronger near-optimal doping compared to the underdoped cuprates.

In the ground state (G-AFM) of YBCO$_6$, the computed magnetic moment per Cu atom is 0.48 $\mu_B$, and the stabilization energy with respect to the non-magnetic phase is 115.9 meV/CuO$_2$. By comparing the energies of the AFM and ferromagnetic (FM) states,(4) we have obtained a value of $J = 125.0$ meV for the magnetic exchange coupling constant, which is in good accord with the corresponding experiment result of $125 \pm 5$ meV in YBCO$_{6.15}$(13). When the various uniform magnetic orders are compared in YBCO$_7$, the G-AFM structure is found to remain the most stable, although with a slightly reduced magnetic moment of 0.42 $\mu_B$/Cu, and a stabilization energy with respect to the non-magnetic phase of 72 meV/CuO$_2$, which is only about 2/3 as large as the corresponding value of 115.9 meV in YBCO$_6$. Since the magnetic moment in the FM state is not sensitive to doping, the computed magnetic coupling $J$ in YBCO$_7$ is reduced to 71.5 meV. Our results explain recent RIXS experiments(*14*) that find a persistence of the Cu magnetic moment with doping in YBCO as well as a decrease in $J$, which is proportional to the initial slope of the spin-wave energy, with increasing doping.

In order to highlight effects of the couplings between the lattice degrees of freedom with the spin and charge degrees of freedom, we considered two exemplar YBCO$_7$ structures for various stripe periodicities where the lattice was kept fixed to that of the G-AFM phase. These are shown as un-relaxed structures in Fig. 1(d) marked with open and filled black circles connected with dashed black lines, which should be compared with the corresponding relaxed structures marked by open and filled red circles connected with solid red lines, respectively. By comparing the relaxed and un-relaxed results, lattice relaxation is seen to play a substantial role in stabilizing the stripe solutions. For example, the lattice relaxation stabilizes the $P_c = 4$ as well as the $P_c = 3$ site-centered stripe state by about 4.5 meV/CuO$_2$ (compare unfilled black and red circles). Along this line, lattice relaxation stabilizes the $P_c = 4$ bond-centered stripe by 2 meV/CuO$_2$ (compare filled black and red circles), and results in replacing the G-AFM phase as the ground state. The $P_c = 4$ bond-centered stripe has been found to be particularly robust in experiments and earlier theoretical studies, especially near 1/8$^{th}$ doping.(*15*) Notably, without lattice relaxation, the $P_c = 4, 6$ and 7 bond-centered stripes are essentially degenerate. Moreover, the trend lines for the two un-relaxed 1D structures in Fig. 1(d) would suggest the most stable stripes to lie near $P_c = 8$, consistent with a QMC study of the one-band Hubbard model of stripes(*12*) in which lattice degrees of freedom were ignored.

Figure 2 presents details of several illustrative stripe structures including lattice distortions resulting from the interplay of spin, charge, and lattice degrees of freedom. The most stable 1D bond-centered and site-centered stripes are considered in panels (a) and (b), respectively, while the 2D stripes are taken up in panels (c) and (d). Stripe formation results in strong modulation of hole doping on the planar oxygen atoms, as seen from the filled green circles at the bottom of panels (a) and (b), with a clear excess of holes in the region of charge stripes where magnetism is reduced or suppressed. In contrast, hole doping on Cu (blue filled squares) is larger on the magnetic stripes, and acts to compensate the oxygen holes. The amplitude of the modulation of holes on planar oxygen sites is larger in the site-centered stripe than the bond-centered case [note differences in vertical scales in panels (a) and (b)]. These findings are in reasonable accord with nuclear magnetic resonance (NMR),(*16*) and scanning-tunneling-microscopy (STM) results,(*17*) and with QMC predictions in a 3-band Hubbard model.(*11*) Our study provides a realistic basis for obtaining the value of the oxygen-to-copper hole ratio, which is a key microscopic quantity that varies systematically over the various cuprate families, and scales with the maximum superconducting T$_c$ for each family.(*18*)

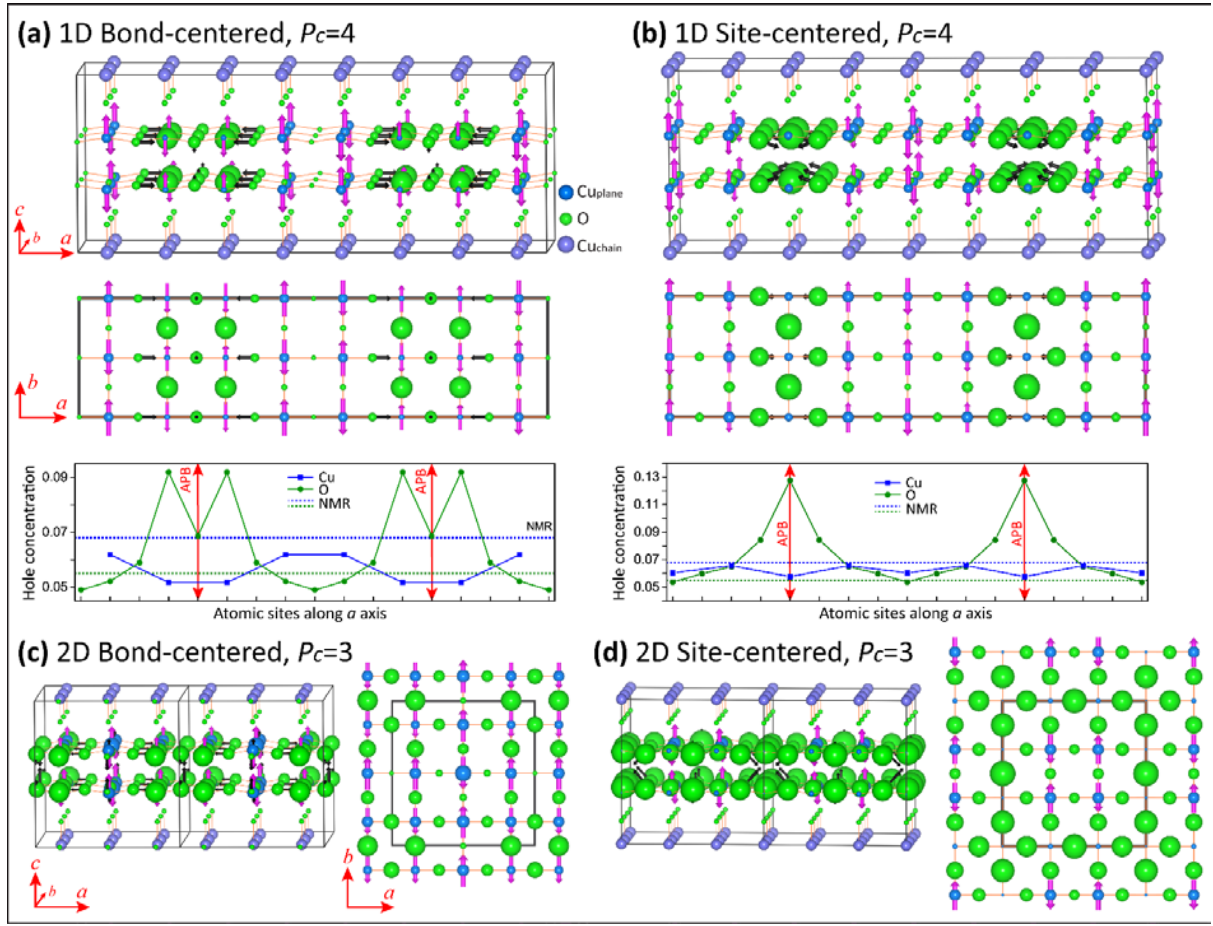

**Fig. 2. Details of modifications resulting in four representative stripe structures in YBCO$_7$ through couplings between spin, charge, and lattice degrees of freedom.** (**a**) A 1D bond-centered stripe perpendicular to the *a*-axis with a periodicity $P_c = 4$. Hole doping on planar Cu and O atoms, defined as the difference in charge between the YBCO$_6$ ground state and the YBCO$_7$ stripe state (see Section 4 in Supplementary Materials for details), is indicated by changes in relative sizes of the atoms. Directions of pink arrows denote spin orientations, and their sizes are proportional to magnetic moments. Black arrows mark directions and sizes of the major atomic displacements resulting from deformation relative to the structure of the G-AFM state of YBCO$_7$. The top panel is a 3D rendition of the structure. Middle panel gives structural details in a CuO$_2$ plane. The bottom panel is a plot of the hole concentration on planar copper atoms and their nearest-neighbor oxygens along the *a*-axis. Y and Ba atoms are omitted in the figure for clarity. The horizontal dashed lines mark the (average) experimental values obtained via NMR.(*16*) (**b**) Same as (a) except that this figure refers to 1D site-centered stripe state. (**c**) 2D bond-centered stripe state with periodicities $P_c = 3$ along both *a*- and *b*-axes. Much of the notation is the same as in panel (a). Black lines mark cell boundaries as well as the APBs. (**d**) Same as (c) except that this figure refers to a 2D site-centered stripe state. See Supplementary Materials for details of other stripe states.

Lattice distortions are particularly prominent around the APBs as we would expect. A reference to black arrows in Fig. 2 shows that the main deformation in both the 1D and 2D stripes involves the movement of oxygen atoms toward the APBs, see Section 3 of Supplementary Materials for details. This results in a Peierls-like distortion of the structure with the appearance of a breathing mode. In the presence of fluctuating stripes, we would then expect the formation of dynamic polarons. The strong spin-charge-lattice coupling indicated by our study suggests that phonons could play a substantial role in creating the superconducting glue in the cuprates.

Interestingly, the 2D stripes in Fig. 2(c,d) are not superpositions of 1D stripes running along two perpendicular directions. Instead, we see what appear to be regular arrays of square magnetic patches framed by APBs. Notably, the $YBCO_7$ orthorhombicity breaks 90º rotational symmetry, more prominently in the bond-centered stripes. The resulting pattern of electronic intensity is in better agreement with STM studies(17) than a crossed-stripe pattern.

What is the underlying microscopic mechanism that drives the material into the stripe order through couplings between the spin, charge, and lattice degrees of freedom? The key is to recognize that when holes are added to the undoped system, they move to disrupt the AFM order. In reaction, there is a tendency for the holes to be expelled from the AFM domains, leaving behind magnetic stripes. The system can then be stabilized naturally by separating the magnetic stripes by APBs and the formation of charge stripes as the excess holes are accommodated in the APBs. The excess holes cause the nearby O ions to shift toward the charge stripes to screen them. The interaction of the lattice with spins and charges thus plays an important role in stabilizing the stripe phases.

In comparing our results with the recent studies of 3-band(11) and 1-band(12) Hubbard model, we should keep in mind that Refs. (11) and (12) neglect lattice relaxation effects and only consider the 1/8$^{th}$ doping case using model dispersions appropriate for $La_{2-x}Sr_xCuO_4$ where enhanced stripe features are expected. Similar to our results, dynamical-mean-field computations in Ref. (11) found the charge modulation on oxygen to be larger than on copper; the failure of their QMC computations to find charge modulation, on the other hand, is not surprising since the computation is at a very high temperature of 970 K. A quasi-degeneracy of stripe energy vs $P_c$ is found in Ref. (12) like in our Fig. 1(d). However, Ref. (12) found a preference for bond-centered stripes for even $P_c$ and site-centered stripes for odd $P_c$, whereas our bond-centered stripes are systematically lower in energy. Also, while $P_c = 4$ is a local minimum, Ref. (12) finds that the ground state corresponds to $P_c = 8$, like the trend in our results neglecting lattice relaxation.

In conclusion, our study demonstrates how competing phases in wide classes of complex quantum materials can be addressed on a first-principles basis without the need to invoke *ad hoc* parameters or to restrict the orbitals included in the underlying Hamiltonian. Our finding that energy minimization is controlled by maximizing the Cu local moment shows that Mott physics continues to play a significant role in YBCO$_7$ and that this physics can be captured effectively within our first-principles framework. Our approach will thus enable a new generation of understanding of the fascinating exotic properties of quantum materials, and how these properties emerge through the interplay of spin, charge, and lattice degrees of freedom.

**COMPUTATIONAL DETAILS**

All calculations were carried out using the Vienna Ab-initio Simulation Package(*19-21*) with the SCAN exchange-correlation functional.(*1,2*) A relatively high cutoff of 500 eV was used to truncate the plane-wave basis set. Section 1 of Supplementary Materials discusses issues related to the convergence of the results with respect to the energy of the plane-wave cutoff and the density of the *k*-mesh; the specific *k*-meshes employed in computations are also given. Spin-orbit coupling effects are neglected in the schematics of Fig. 1(a,b), and the spins are shown as vertical arrows for simplicity. For the uniform magnetic phases, spin-orbit coupling causes the spins to lie mostly in-plane, see Section 2 of Supplementary Materials, which is consistent with experimental findings and our calculations on $La_{2-x}Sr_xCuO_4$.(*4*) All crystal structures were fully relaxed with a force convergence criterion of 0.01 eV/Å, unless specified otherwise. Stripes are introduced into superlattices by assigning the spin directions on the planar Cu ions with a specific starting pattern, but then all spin, charge, and lattice degrees of freedom were allowed to relax.

**Acknowledgments:** We thank Hoang Tran, Carl Baribault, and Hideki Fujioka for their computational support at Tulane University. The work at Tulane University was supported by the start-up funding from Tulane University, the Cypress Computational Cluster at Tulane, the DOE Energy Frontier Research Centers (development and applications of density functional theory): Center for the Computational Design of Functional Layered Materials (DE-SC0012575), and the National Energy Research Scientific Computing Center supercomputing center (DOE grant number DEAC02-05CH11231). The work at Northeastern University was supported by the U.S. DOE, Office of Science, Basic Energy Sciences grant number DE-FG02-07ER46352 (core research) and benefited from Northeastern University's Advanced Scientific Computation Center, the National Energy Research Scientific Computing Center supercomputing center (DOE grant number DEAC02-05CH11231), and support (testing the efficacy of new functionals in diversely bonded materials) from the DOE Energy Frontier Research Centers: Center for the Computational Design of Functional Layered Materials (DE-SC0012575).

**Author contributions:** YZ, CL, and JWF performed computations and analyzed the data. BB, RSM, AB, and JS led the investigations, designed the computational approaches, provided computational infrastructure and analyzed the results. All authors contributed to the writing of the manuscript;

**Competing interests:** The authors declare no competing interests.

**Data and materials availability:** All data is available in the main text or the supplementary materials.